\documentclass[showpacs,preprintnumbers,amsmath,amssymb]{revtex4}
\usepackage{amsmath,amsfonts,amssymb}
\usepackage{epsfig}
\def\bit{\begin{itemize}}
\def\eit{\end{itemize}}
\def\ben{\begin{enumerate}}
\def\een{\end{enumerate}}
\def\bed{\begin{description}}
\def\eed{\end{description}}

\def\lsim{\raise0.3ex\hbox{$<$\kern-0.75em\raise-1.1ex\hbox{$\sim$}}}
\def\gsim{\raise0.3ex\hbox{$>$\kern-0.75em\raise-1.1ex\hbox{$\sim$}}}

\let\jnfont=\rm
\def\NPB#1,{{\jnfont Nucl.\ Phys.\ B }{\bf #1},}
\def\PLB#1,{{\jnfont Phys.\ Lett.\ B }{\bf #1},}
\def\EPJC#1,{{\jnfont Eur.\ Phys.\ Jour.\ C }{\bf #1},}
\def\PRD#1,{{\jnfont Phys.\ Rev.\ D }{\bf #1},}
\def\PRL#1,{{\jnfont Phys.\ Rev.\ Lett.\ }{\bf #1},}
\def\MPLA#1,{{\jnfont Mod.\ Phys.\ Lett.\ A }{\bf #1},}
\def\JPG#1,{{\jnfont J.\ Phys.\ G}{\bf #1},}
\def\CTP#1,{{\jnfont Commun.\ Theor.\ Phys.\ }{\bf #1},}
\def\JHEP#1,{{\jnfont JHEP \ }{\bf #1},}
\def\NPPS#1,{{\jnfont Nucl.\ Phys.\ Proc.\ Suppl.\ }{\bf #1},}

\def\beq{\begin{equation}}
\def\eeq{\end{equation}}
\def\bea{\begin{eqnarray}}
\def\eea{\end{eqnarray}}
\newcommand{\ba}{\begin{array}}
\newcommand{\ea}{\end{array}}

\begin{document}

\title{SUSY dark matter in light of CDMS II results:  a comparative study \\
       for different models }

\author{\ \\[1mm]
Junjie Cao$^1$, Ken-ichi Hikasa$^2$, Wenyu Wang$^3$, Jin Min Yang$^4$, Li-Xin Yu$^4$ \\ ~
}

\affiliation{
$^1$ Department of Physics, Henan Normal university, Xinxiang 453007, China\\
$^2$ Department of Physics, Tohoku University, Sendai 980-8578, Japan \\
$^3$ Institute of Theoretical Physics, College of Applied Science,
     Beijing University of Technology, Beijing 100124, China\\
$^4$ Key Laboratory of Frontiers in Theoretical Physics,
     Institute of Theoretical Physics, Academia Sinica,
              Beijing 100190, China}

\begin{abstract}
We perform a comparative study of  the neutralino dark matter
scattering on nucleon in three popular supersymmetric models: the
minimal (MSSM), the next-to-minimal (NMSSM) and the nearly minimal
(nMSSM). First, we give the predictions of the elastic cross section
by scanning over the parameter space allowed by various direct and
indirect constraints, which are from the measurement of the cosmic
dark matter relic density, the collider search for Higgs boson and
sparticles, the precision electroweak measurements and the muon
anomalous magnetic moment. Then we demonstrate the property of the
allowed parameter space with/without the new limits from CDMS II. We
obtain the following observations: (i) For each model the new CDMS
limits can exclude a large part of the parameter space allowed by
current collider constraints; (ii) The property of the allowed
parameter space is similar for MSSM and NMSSM, but quite different
for nMSSM; (iii) The future SuperCDMS can cover most part of the
allowed parameter space for each model.
\end{abstract}
\maketitle

\section{Introduction}
Although there are many theoretical or aesthetical arguments for
the necessity of TeV-scale new physics, the most convincing
evidence is from the WMAP (Wilkinson Microwave Anisotropy Probe)
observation of the cosmic cold dark matter, which naturally
indicate the existence of WIMPs (Weakly Interacting Massive
Particle) beyond the prediction of the Standard Model (SM).  By
contrast, the neutrino oscillation may rather imply trivial new
physics (plainly adding right-handed neutrinos to the SM) or new
physics at some very high see-saw scale unaccessible to any
foreseeable colliders. Therefore, the TeV-scale new physics to be
unravelled at the Large Hadron Collider (LHC) is most likely
related to the WIMP dark matter.

If WIMP dark matter is chosen by nature, then it will naturally
direct to low-energy supersymmetry (SUSY) with R-parity although
other miscellaneous speculations are also possible. In addition to
the perfect explanation of cosmic dark matter, to make perfection
still more perfect, SUSY can also solve another plausible puzzle,
namely the $3\sigma$ deviation of the muon anomalous magnetic
moment from the SM prediction. In the framework of SUSY, the most
intensively studied model is the minimal supersymmetric standard
model (MSSM) \cite{mssm}, which is the most economical realization
of SUSY. Since this model suffers from the $\mu$-problem and the
little hierarchy problem, other supersymmetric models have
recently attracted much attention, among which is the extension by
introducing a gauge singlet superfield $\hat{S}$, such as the
next-to-minimal supersymmetric model (NMSSM) \cite{NMSSM} and the
nearly minimal supersymmetric model (nMSSM)
\cite{xnMSSM1,xnMSSM2}. In addition to the attractive
phenomenological virtues like the alleviation of the little
hierarchy problem and the possible explanation \cite{hooper} of
PAMELA positron excess (albeit subject to large uncertainty and
could be explained astrophysically by pulsars) \cite{pamela}, such
singlet extensions are arguably motivated by some fancy string
theory, e.g., the NMSSM can be constructed from a heterotic string
\cite{string}.

In this work, motivated by the CDMS II new results
\cite{cdms2,cdms2-a}, we examine the SUSY dark matter scattering on
the nucleon ($\chi$-nucleon scattering). In the literature such a
topic has been studied mainly in the constrained MSSM \cite{ellis}.
Our work is projected to have the following features:
\begin{itemize}
\item[(i)]  We perform a comparative study for three popular SUSY models:
the MSSM, the NMSSM and the nMSSM.
\item[(ii)] We consider the constraints from the cosmic dark matter relic density
and current collider experiments, such as
the collider search for Higgs boson and sparticles,
the precision electroweak
measurements and the muon anomalous magnetic moment.
By scanning over the
parameter space subject to these constraints,
 for each model we find out the allowed parameter space
and give the predictions of the cross section for $\chi$-nucleon
scattering with comparison to the CDMS II results. \item[(iii)] We
demonstrate the properties of the allowed parameter space (such as
the components of the neutralino dark matter and the invisible
Higgs boson decay into a pair of dark matter particles) by
comparing the three models. \item[(iv)] We show the capability of
the SuperCDMS \cite{supercdms} in probing the currently allowed parameter space for
each model.
\end{itemize}

This paper is organized as follows. In Sec.II we briefly describe
the three models: the MSSM, the NMSSM and the nMSSM, focusing on
the Higgs sector and the neutralino/chargino sector since they are
directly relevant to the dark matter scattering. In Sec.III we
scan over the parameter space under current constraints, and give
the predictions of the cross section for $\chi$-nucleon scattering
with comparison to the CDMS II results. Also we will demonstrate
the properties of the allowed parameter space with/without
considering the CDMS new limits. In Sec. IV we give our
conclusions.

\section{Supersymmetric models}
\label{sec2}

As the economical realizations of supersymmetry, the MSSM has the
minimal content of particles, while the NMSSM and nMSSM extend the
MSSM by only adding one singlet Higgs superfield $\hat{S}$. The
difference between these models is reflected in their
superpotential:
\begin{eqnarray}
W_{\rm MSSM}& = & W_F+\mu \hat H_u\cdot\hat H_d,\\
W_{\rm NMSSM} & = & W_F + \lambda \hat{H}_u\cdot\hat{H}_d \hat{S}
                  +\frac{1}{3} \kappa  \hat{S}^3, \\
W_{\rm nMSSM} & = &  W_F + \lambda \hat{H}_u\cdot\hat{H}_d \hat{S}
                  + \xi_F M_n^2 \hat{S},
\label{superpotential}
\end{eqnarray}
where
$W_F= Y_u  \hat{Q}\cdot\hat{H}_u  \hat{U}
    -Y_d \hat{Q}\cdot\hat{H}_d \hat{D}
    -Y_e \hat{L}\cdot\hat{H}_d \hat{E}$
with $\hat{Q}$, $\hat{U}$ and $\hat{D}$ being the squark
superfields, and $\hat{L}$ and $\hat{E}$ being the slepton
superfields, $\hat{H}_u$ and $\hat{H}_d$ are the Higgs doublet
superfields,  $\lambda$, $\kappa$ and $\xi_F$ are dimensionless
coefficients, and $\mu$ and $M_n$ are parameters with mass
dimension. Note that there is no explicit $\mu$-term in the NMSSM or
nMSSM, and an effective $\mu$-parameter can be generated when the
scalar component ($S$) of $\hat{S}$ develops a vev (vacuum
expectation value). Also note that the nMSSM differs from the NMSSM
in the last term with the trilinear singlet term $\kappa \hat{S}^3$
of the NMSSM replaced by the tadpole term $\xi_F M_n^2 \hat{S}$. As
pointed out in \cite{xnMSSM1}, such a tadpole term can be generated
at a high loop level and naturally be of the SUSY breaking scale.
The advantage of such replacement is the nMSSM has no discrete
symmetry and thus free of the domain wall problem which the NMSSM¡¡
suffers  from.

Corresponding to the superpotential, the Higgs soft terms in the
scalar potentials are also different for three models (the soft
terms for gauginos and sfermions are the same and not listed here)
 \begin{eqnarray}
V_{\rm soft}^{\rm MSSM}&=&\tilde m_{d}^{2}|H_d|^2 + \tilde m_{u}^{2}|H_u|^2
            + \left( B\mu H_u\cdot H_d +  \mbox{h.c.} \right) \\
V_{\rm soft}^{\rm NMSSM}&=&\tilde m_{d}^{2}|H_d|^2 + \tilde m_{u}^{2}|H_u|^2
            + \tilde m_s^{2}|S|^2
           + \left( A_\lambda \lambda S H_d\cdot H_u
           + \frac{\kappa}{3} A_{\kappa} S^3 + \mbox{h.c.} \right) , \\
V_{\rm soft}^{\rm nMSSM} & = & \tilde{m}_d^2 |H_d|^2 + \tilde{m}_u^2 |H_u|^2
   + \tilde{m}_s^2 |S|^2
   + \left( A_\lambda \lambda S H_d\cdot H_u + \xi_S M_n^3 S
   + \mbox{h.c.} \right).
\label{soft}
\end{eqnarray}
After the scalar fields $H_u$,$H_d$ and $S$ develop their vevs
$v_u$, $v_d$ and $s$ respectively, they can be expanded as
\begin{eqnarray}
H_d  = \left ( \begin{array}{c}
             \frac{1}{\sqrt{2}} \left( v_d + \phi_d + i \varphi_d \right) \\
             H_d^- \end{array} \right) \, ,
H_u  = \left ( \begin{array}{c} H_u^+ \\
       \frac{1}{\sqrt{2}} \left( v_u + \phi_u + i \varphi_u \right)
        \end{array} \right)  \, ,
S  = \frac{1}{\sqrt{2}} \left( s + \sigma + i \xi \right)  \, .
\end{eqnarray}
The mass eigenstates can be obtained by unitary rotations
\begin{eqnarray}
\left( \begin{array}{c} H_1 \\ H_2\\ H_3\end{array} \right)
= U^H \left( \begin{array}{c} \phi_d \\ \phi_u \\ \sigma \end{array} \right),~
\left(\begin{array}{c} A_1 \\ A_2 \\ G_0\end{array} \right)
= U^A \left(\begin{array}{c}\varphi_d \\ \varphi_u\\ \xi \end{array} \right),~
 \left(\begin{array}{c}G^+\\ H^+ \end{array} \right)
=U^{H^+} \left(\begin{array}{c} H_d^+ \\ H_u^+\end{array}  \right),
\end{eqnarray}
where $H_{1,2,3}$ and $A_{1,2}$ are respectively the CP-even and CP-odd
neutral Higgs bosons, $G^0$ and $G^+$ are Goldstone bosons, and $H^+$ is
the charged Higgs boson. So in the NMSSM and nMSSM, there exist
a pair of charged Higgs bosons, three
CP-even and two CP-odd neutral Higgs bosons.
In the MSSM, due to the absence of $S$, we only have two CP-even and one CP-odd
neutral Higgs bosons in addition to a pair of charged Higgs bosons.

The MSSM predict four neutralinos  $\chi^0_i$ ($i=1,2,3,4$), i.e.
the mixture of neutral gauginos (bino $\lambda'$ and neutral wino
$\lambda^3$) and neutral Higgsinos ($\psi_{H_u}^0, \psi_{H_d}^0$),
while the NMSSM and nMSSM predict one more neutralino because the
singlino $\psi_S$ comes into the mixing. In the basis $(-i\lambda',
- i \lambda^3, \psi_{H_u}^0, \psi_{H_d}^0, \psi_S )$ (for MSSM
$\psi_S$ is absent) the  neutralino mass matrix is given by
\begin{eqnarray}\label{mass-matrix1}
&
\left( \begin{array}{cccc}
M_1          & 0             & m_Zs_W s_b    & - m_Z s_W c_b   \\
0            & M_2           & -m_Z c_W s_b  & m_Z c_W c_b     \\
m_Zs_W s_b   & -m_Z s_W s_b  & 0             & -\mu            \\
-m_Z s_W c_b & -m_Z c_W c_b  &  -\mu         & 0               \\
\end{array} \right) ~~~~~~~~& {\rm ~~for ~MSSM} \\
&
\left( \begin{array}{ccccc}
M_1          & 0             & m_Zs_W s_b    & - m_Z s_W c_b  & 0 \\
0            & M_2           & -m_Z c_W s_b  & m_Z c_W c_b    & 0 \\
m_Zs_W s_b   & -m_Z s_W s_b  & 0             & -\mu           & -\lambda v c_b \\
-m_Z s_W c_b & -m_Z c_W c_b  &  -\mu         & 0              & - \lambda v s_b \\
0            & 0             &-\lambda v c_b &- \lambda v s_b & 2 \frac{\kappa}{\lambda}\mu
\end{array} \right) & {\rm ~~for ~NMSSM} \\
&
\left( \begin{array}{ccccc}
M_1          & 0             & m_Zs_W s_b    & - m_Z s_W c_b  & 0 \\
0            & M_2           & -m_Z c_W s_b  & m_Z c_W c_b    & 0 \\
m_Zs_W s_b   & -m_Z s_W s_b  & 0             & -\mu           & -\lambda v c_b \\
-m_Z s_W c_b & -m_Z c_W c_b  &  -\mu         & 0              & - \lambda v s_b \\
0            & 0             &-\lambda v c_b &- \lambda v s_b & 0
\end{array} \right) & {\rm ~~for ~nMSSM}
\label{mass-matrix3}
\end{eqnarray}
where $M_1$ and $M_2$ are respectively $U(1)$ and $SU(2)$ gaugino masses,
$s_W=\sin \theta_W$, $c_W=\cos\theta_W$, $s_b=\sin\beta$ and $c_b=\cos\beta$
with $\tan \beta \equiv v_u/v_d$.
In our study the lightest neutralino  $\chi^0_1$ is assumed to
be the lightest supersymmetric particle (LSP), serving as the SUSY dark matter
particle. It is composed by
\begin{eqnarray}
\chi^0_1=N_{11}(-i\lambda_1)+N_{12}(-i \lambda_2)
+N_{13}\psi_{H_u}^0 +N_{14} \psi_{H_d}^0 + N_{15}\psi_S,
\end{eqnarray}
where $N$ is the unitary matrix ($N_{15}$ is zero for the MSSM)
to diagonalize the mass matrix in
Eqs.(\ref{mass-matrix1}-\ref{mass-matrix3}).

The chargino sector of these three models is the same except
that for the NMSSM/nMSSM the parameter $\mu$ is replaced by $\mu_{\rm eff}$.
The charginos $\chi^\pm_{1,2}$ ($m_{\chi^\pm_1}\leq m_{\chi^\pm_2}$)
are the mixture of charged Higgsinos $\psi_{H_{u,d}}^\pm$
and winos $\lambda^\pm=(\lambda^1\pm\lambda^2)/\sqrt{2}$, whose mass
matrix in the basis of  $(-i\lambda^\pm,\psi_{H_{u,d}}^\pm)$ is given by
\begin{eqnarray}
\left( \begin{array}{cc}
 M_2 &             \sqrt{2} m_W\sin\beta \\
\sqrt{2} m_W\cos\beta & \mu_{\rm eff}
\end{array} \right).
\end{eqnarray}
So the chargino $\chi^\pm_1$ can be wino-dominant (when $M_2$ is much smaller than
$\mu$) or Higgsino-dominant (when $\mu$ is much smaller than $M_2$).
Since the composing property (wino-like, bino-like, Higgsino-like or singlino-like)
of the LSP and the chargino $\chi^\pm_1$ is very important for SUSY phenomenology,
we will show such a property in our following study.

\section{numerical results and discussions}
\label{sec3}
So far there are various constraints from both collider and dark matter experiments.
In our study we consider the following constraints:
\begin{itemize}
\item[(1)] Direct bounds on sparticle and Higgs masses from LEP and Tevatron
experiments \cite{Yao},
e.g.,$m_{\chi^+_1} > 103.5$ GeV, $m_{\tilde{e}} > 73  {\rm ~GeV}$,
$m_{\tilde{\mu}} > 94  {\rm~GeV}$,
$m_{\tilde{\tau}} > 81.9$ GeV and $m_{H^+} > 78.6$ GeV.
\item[(2)] LEP II search for Higgs boson \cite{Higgs}, which include
various channels of Higgs boson productions \cite{NMSSMTools}.
\item[(3)] LEP I and LEP II constraints on the productions of neutralinos and charginos,
including the LEP I invisible Z-decay $\Gamma(Z \to \chi^0_1 \chi^0_1) < 1.76$ MeV,
the LEP II neutralino production $\sigma(e^+e^- \to \chi^0_1 \chi^0_i) < 10^{-2}~{\rm pb}$ ($i>1$)
and $\sigma(e^+e^- \to \chi^0_i \chi^0_j) < 10^{-1}~{\rm pb}$.
\item[(4)] Indirect constraints from precision electroweak observables such
as $\rho_{\ell}$, $\sin^2 \theta_{eff}^{\ell}$ and $M_W$, or their
combinations $\epsilon_i (i=1,2,3)$ \cite{Altarelli}. We require
$\epsilon_i$ to be compatible with the LEP/SLD data at $95\%$ confidence level.
Also, for $R_b = \Gamma (Z \to \bar{b} b) / \Gamma
( Z \to {\rm hadrons} )$ whose measured value is $R_b^{exp} =
0.21629 \pm 0.00066 $ and the SM prediction is $R_b^{SM} = 0.21578 $
for $m_t = 173$ GeV \cite{Yao}, we require
$R_b^{SUSY}$ is within the $2 \sigma$ range of its experimental value.
Various B-physics constraints are also included  \cite{NMSSMTools}.
\item[(5)] Indirect constraint from the muon anomalous magnetic moment,
 $a_\mu^{exp} - a_\mu^{SM} = ( 25.5 \pm 8.0 ) \times 10^{-10} $ \cite{Davier},
for which we require the SUSY effects to account at $2 \sigma$ level.
(We note that $3 \sigma$ effects are considered to be inconclusive in high 
energy physics.  In collider experiments, there are a large number of channels 
and observables and there is a good chance that some of the measurements 
can show such deviation from expectation.  The muon $g-2$ experiment 
is quite different because there is just one quantity to measure 
in the experiment. In our opinion, the significance of the deviation 
should be taken rather seriously.)  
\item[(6)] Dark matter constraints from the WMAP relic density
$ 0.0945 < \Omega h^2 < 0.1287 $ \cite{dmconstr}
and CDMS II limits on the scattering cross section \cite{cdms2}.
To show the effects of the CDMS II limits, we will display the results
with/without such limits.
\end{itemize}
In addition to the above experimental limits, we also consider the
constraint from the stability of the Higgs potential, which requires
that the physical vacuum of the Higgs potential with non-vanishing
vevs of Higgs scalars should be lower than any local minima.
Further, the soft breaking parameters are required to be below 1 TeV to
avoid the fine-tuning, and $\lambda$ (at weak scale) is less than
about 0.7 to ensure perturbativity of the theory up to  the grand
unification scale ($\lambda$ is increasing with the energy scale
\cite{zerwas}). Note that most of these constraints have been
encoded in NMSSMTools \cite{NMSSMTools}. We extend this package and
use it in our calculations. For the cross section of $\chi$-nucleon
scattering, we use the formulas in \cite{Drees,susy-dm-review} for
the MSSM and extend them to the NMSSM/nMSSM (see Appendix A).

Considering all the constraints listed above,
we scan over the parameters in the following ranges
\begin{eqnarray}
&& 100 {\rm ~GeV} \leq
\left(M_{\rm soft}^{\rm squark},M_{\rm soft}^{\rm slepton}, ~m_A, ~\mu \right) \leq  1 {\rm ~TeV},
\nonumber\\
&& 50 {\rm ~GeV} \leq M_1 \leq  1 {\rm ~TeV}, ~~1 \leq  \tan \beta \leq 40, \nonumber\\
&& \left( |\lambda|,|\kappa| \right) \leq 0.7, ~~|A_\kappa| \leq  1 {\rm ~TeV},
\end{eqnarray}
To reduce the number of the relevant soft parameters, we work in the
so-called $m_h^{max}$ scenario with following choice of the soft
masses for the third generation squarks:
$M_{Q_3}=M_{U_3}=M_{D_3}=800$ GeV, and $X_t = A_t - \mu \cot \beta =
-1600$ GeV.  The advantage of such a choice is that other SUSY
parameters are easy to survive the constraints (so that the bounds
we obtain are conservative). Moreover, we assume
the grand unification relation for the gaugino masses: $M_1:M_2:M_3
\simeq 1:1.83:5.26$ and also assume universal masses $M_{\tilde{\ell}}$
and $M_{\tilde{q}}$  for the three generations of sleptons and the
first two generations of squarks respectively.

 \begin{figure}[htbp]
 \epsfig{file=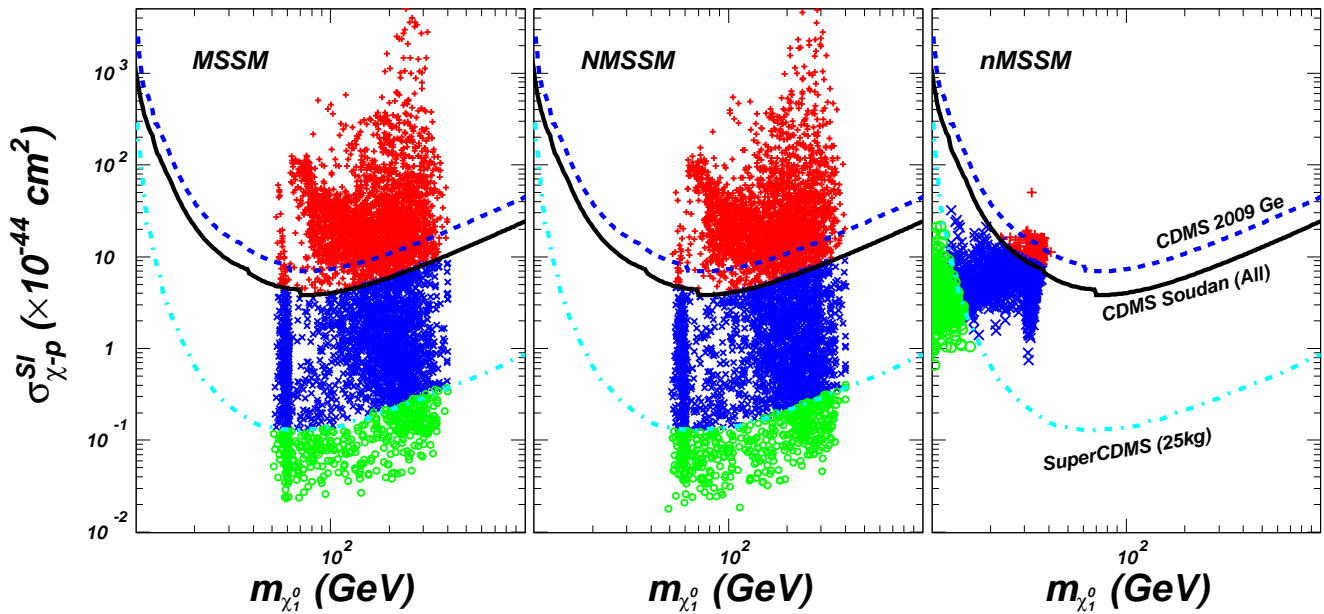,height=8.5cm}
 \vspace{-0.3cm}
\caption{ The scatter plots for the spin-independent elastic cross
section of $\chi$-nucleon scattering. The `$+$' points (red) are excluded
          by CDMS limits (solid line),  the `$\times$' (blue) would be further excluded by SuperCDMS 25kg 
          \cite{supercdms}
          in case
          of unobservation (dash-dotted line), and the `$\circ$' (green) are beyond the SuperCDMS sensitivity.}
   \label{fig1}
    \end{figure}
The surviving points are displayed in Fig.~\ref{fig1} for the
spin-independent elastic cross section of $\chi$-nucleon scattering.
We see that for each model the CDMS II limits can exclude a large
part of the parameter space allowed by current collider constraints
and the future SuperCDMS (25 kg) can cover the most part of the
allowed parameter space. For the MSSM and NMSSM the dark matter mass
range $m_{\chi_1^0}$ is from 50 GeV to 400 GeV, while for the nMSSM
the dark matter mass is constrained below 40 GeV by current
experiments and further constrained below 20GeV by SuperCDMS in case
of unobservation. For the MSSM/NMSSM the LSP lower bound at 50 GeV 
is from the chargino lower bound of 103.5 GeV plus the assumed 
GUT relation $M_1 \simeq 0.5 M_2$; while the upper bound at 400 GeV 
is from the bino nature of the LSP ($M_1$ cannot be too large, 
must be much smaller than other relevant parameters) plus the constraints 
from the LEP II search for Higgs bosons, the muon g-2 and B-physics.
Note that if we do not assume the GUT relation $M_1 \simeq 0.5 M_2$, 
then $M_1$ can be as small as 40 GeV and the LSP lower bound in MSSM/NMSSM
will not be sharply at 50 GeV.

\begin{figure}[htbp]
 \epsfig{file=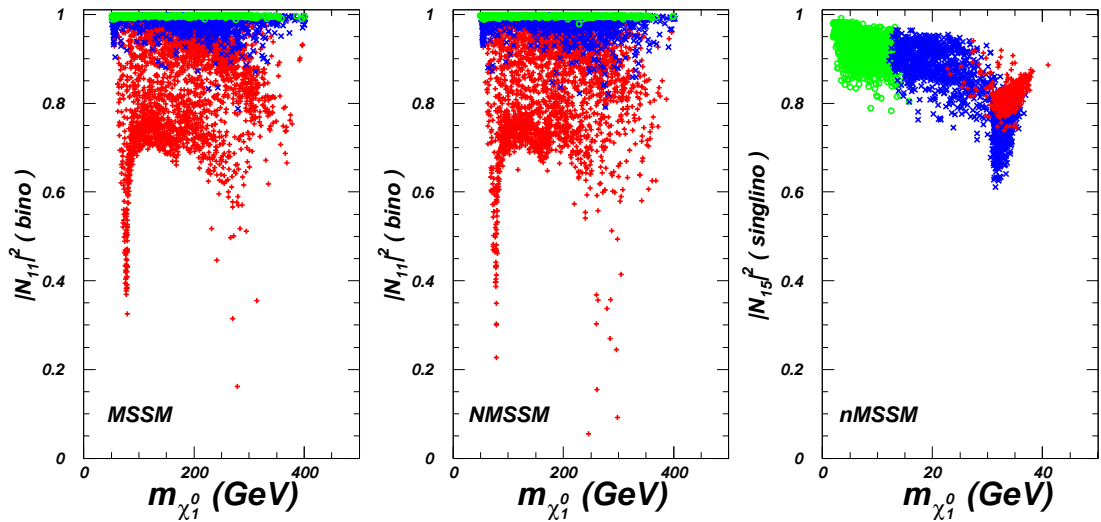,height=7cm}
 \vspace{-0.3cm}
\caption{Same as Fig.~\ref{fig1}, but projected on the plane of
         $|N_{11}|^2$ and $|N_{15}|^2$ versus dark matter mass.}
\label{fig2}
\end{figure}
In Fig.~\ref{fig2} we show the bino component of $\chi_1^0$ in
MSSM/NMSSM and the singlino component of $\chi_1^0$ in nMSSM. We see
that for both the MSSM and NMSSM $\chi_1^0$ is bino-dominant, while
for the nMSSM $\chi_1^0$ is singlino-dominant, and  the region
allowed by CDMS limits (and SuperCDMS limits in case of
unobservation) favor a more bino-like $\chi_1^0$ for the  MSSM/NMSSM
and a more singlino-like $\chi_1^0$ for the nMSSM. For the
MSSM/NMSSM, the reason is obvious because the dominant contribution
to the cross section comes from Fig.~\ref{Feyn1} in the Appendix and
a more bino-like $\chi_1^0$ tends to suppress not only
$f_{q_i}^{\tilde{q}}$ in Eq.(\ref{treecontr}) \cite{Drees}, but also
$f_{q_i}^{H}$ by diminishing $T_{h00}$. As for the nMSSM, $\chi_1^0$
is singlino-like due to the small singlino mass in the
neutralino mass matrix. The peculiarity of the nMSSM
predictions will be discussed at the end of this
section.

\begin{figure}[htbp]
 \epsfig{file=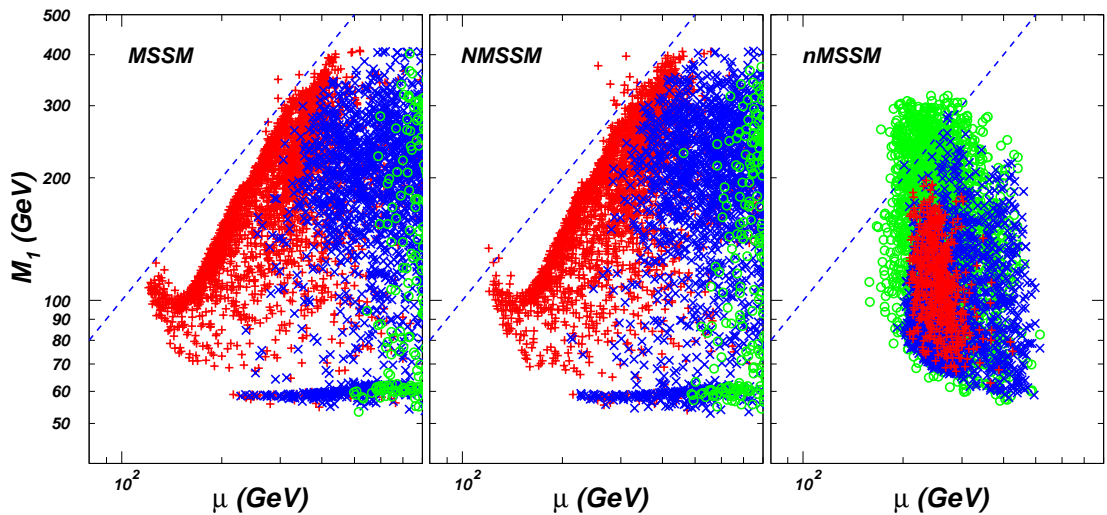,height=7cm}
\vspace{-0.3cm}
\caption{Same as Fig.~\ref{fig1}, but projected on the plane of
         $M_1$ versus $\mu$. The dashed lines are for $M_1=\mu$.}
\label{fig3}
\end{figure}

\begin{figure}[htbp]
\epsfig{file=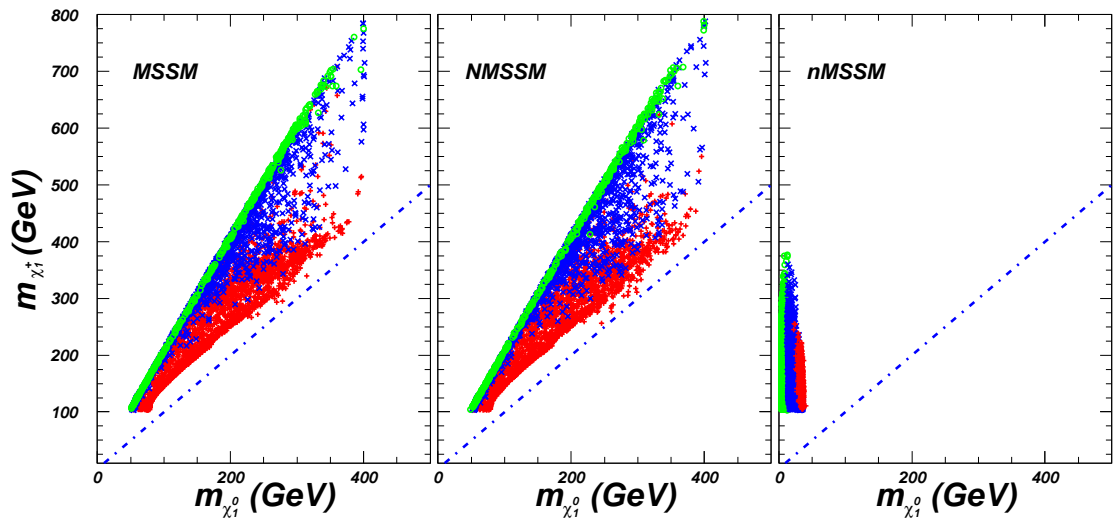,height=7cm} \vspace{-0.3cm} \caption{Same as
Fig.~\ref{fig1}, but showing
         the chargino mass $m_{\chi^+_1}$ versus the LSP mass.
         The dashed lines indicate $m_{\chi^+_1}=m_{\chi^0_1}$.}
\label{fig4}
\end{figure}

In Fig.~\ref{fig3} we project the surviving points on the plane of
$M_1$ versus $\mu$. We see that for both the MSSM and NMSSM most of the
survived points are below the $M_1=\mu$ line, implying that
$\chi_1^0$ is bino-dominant. The region allowed by the CDMS limits
tends to have a larger $\mu$, indicating a more bino-like $\chi_1^0$,
which can be inferred from the neutralino mass matrices in
Eq.(\ref{mass-matrix3}).
For the nMSSM the upper bound of 500 GeV for $\mu$ is from the
fact that a larger $\mu$ leads to a lighter LSP (as will be shown in 
Eq.\ref{m_chi0}),
which is then constrained by the required annihilation rate of the LSP.

In Fig.~\ref{fig4} we display the surviving points on the plane of
the chargino mass $m_{\chi^+_1}$ versus $m_{\chi^0_1}$. For both
the MSSM and NMSSM, the CDMS limits tend to favor a heavier
chargino and ultimately the SuperCDMS limits tend to favor a
wino-dominant chargino with mass about $2 m_{\chi^0_1}$. This can
be understood because the CDMS/SuperCDMS limits require a large
$\mu$, which makes $\chi^+_1$ to be dominated by wino with a mass
$M_2\simeq 2M_1$.

\begin{figure}[htbp]
\epsfig{file=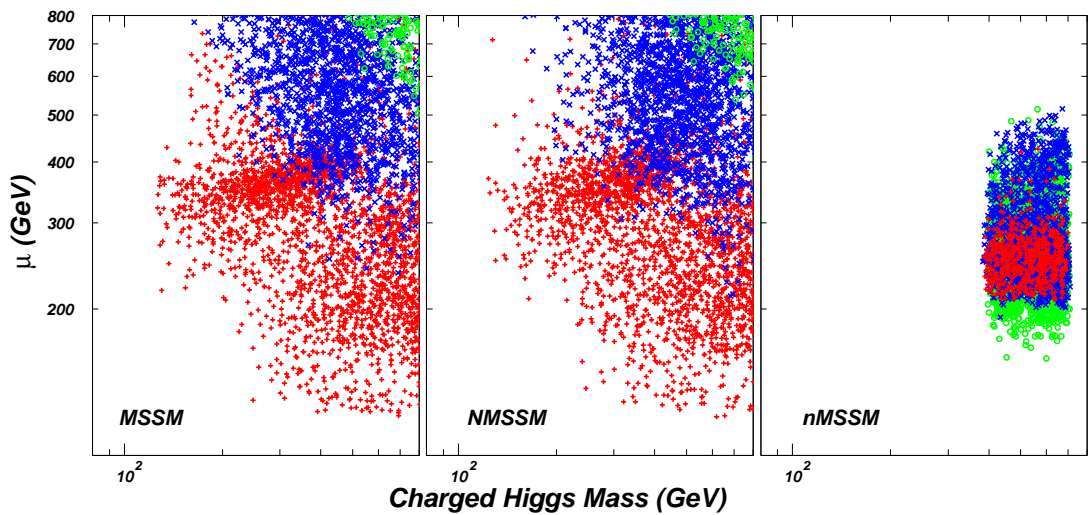,height=7cm}
\vspace{-0.3cm}
\caption{Same as Fig.~\ref{fig1}, but projected on the plane of
         $\mu$ versus the charged Higgs mass.}
\label{fig5}
    \end{figure}
\begin{figure}[htbp]
\epsfig{file=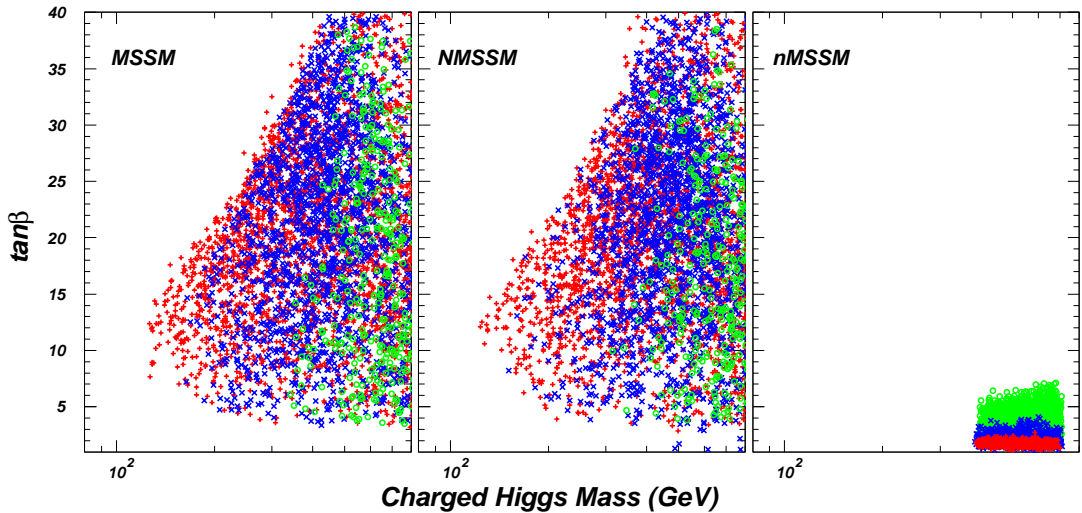,height=7cm}
\vspace{-0.3cm}
\caption{Same as Fig.~\ref{fig1}, but projected on the plane of
         $\tan\beta$ versus the charged Higgs mass.}
\label{fig6}
    \end{figure}

\begin{figure}[htbp]
\epsfig{file=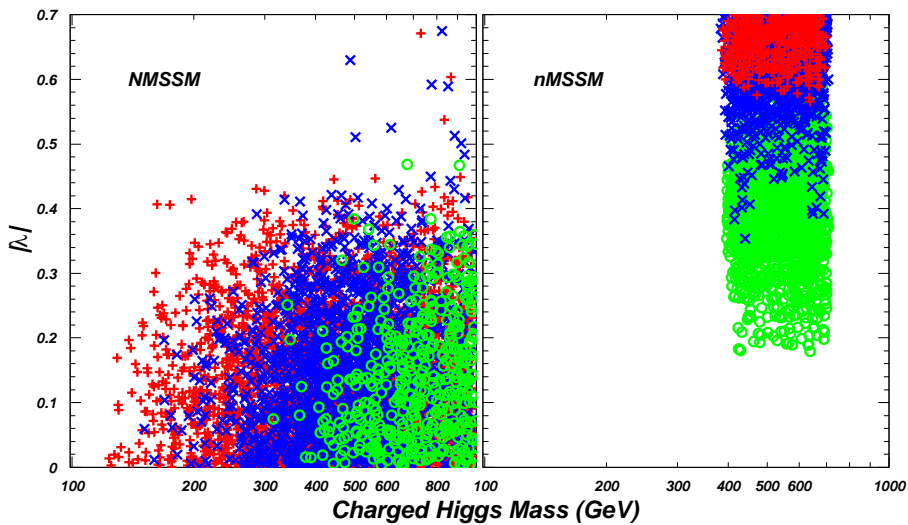,height=7cm} \vspace{-0.3cm} \caption{Same as
Fig.~\ref{fig1}, but projected on the plane of
         $|\lambda|$ versus the charged Higgs mass in NMSSM and nMSSM.}
\label{fig7}
    \end{figure}

\begin{figure}[htbp]
\epsfig{file=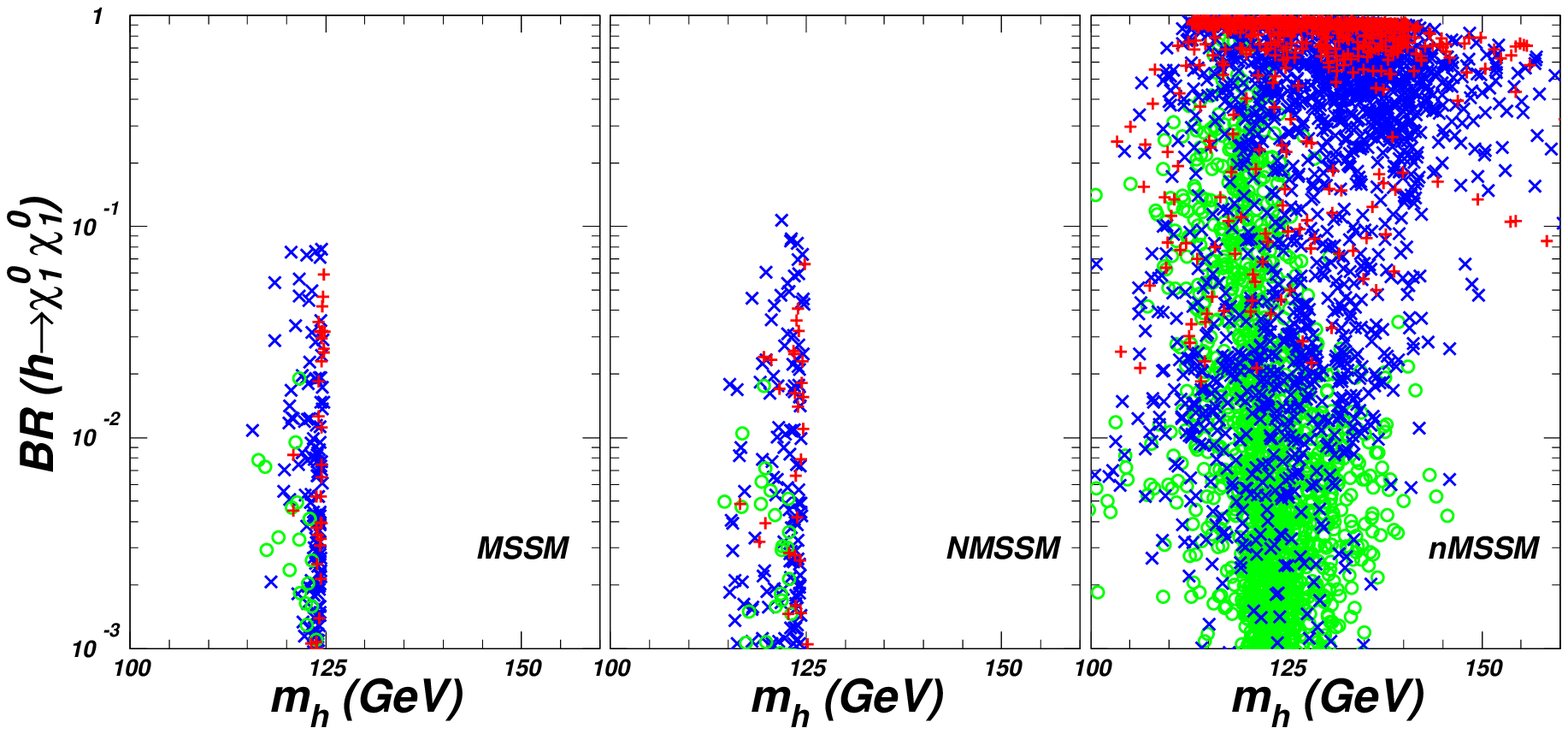,height=7cm} \vspace{-0.3cm} \caption{Same as
Fig.~\ref{fig1}, but projected for
         the decay branching ratio of $h^0 \to \chi^0_1 \chi^0_1$
         versus the mass of the Higgs boson $h$ (the SM-like Higgs boson).}
\label{fig8}
    \end{figure}

In Figs.\ref{fig5} and \ref{fig6} we display the surviving points on
the plane of $\mu$ and $\tan\beta$ versus the charged Higgs mass. In
both the MSSM and NMSSM, large $\mu$ and small $\tan \beta$ are
favored for a light charged Higgs boson. The reason is as follows.
In the MSSM, there are two CP-even Higgs bosons contributing to the
cross section. One is the SM-like Higgs boson $h^0$ with mass around
$120$GeV and the other is the heavy boson $H^0$ with mass nearly
degenerate with the charged Higgs boson. Then from the expression
of $f_{q_i}^H$ in Eq.(\ref{treecontr}), one can learn that the $H^0$
contribution to the scattering cross section get enhanced for light
charged Higgs boson. In this case, to alleviate such enhancement,
large $\mu$ (to lower $T_{H00}$) and/or small $\tan \beta$ (to lower
$T_{hq_iq_i}$) are needed. In the NMSSM, although there are three
CP-even Higgs boson contributing to the scattering, we can get the
same conclusion as the MSSM because one of the bosons is
singlet-dominant and its contribution is suppressed by
$T_{hq_iq_i}$, and the contributions from the other two bosons are
quite similar to the case of the MSSM, .

In Fig.~\ref{fig7} we show the value of $|\lambda|$ versus the
charged Higgs mass in NMSSM and nMSSM. This figure indicates that
$\lambda$ larger than 0.4 is disfavored for the NMSSM. The
underlying reason is that $T_{h00}$ in Eq.(\ref{treecontr}) depends on
$\lambda$ explicitly and large $\lambda$ can enhance $T_{h00}$
\cite{NMSSMTools}.  By contrast, although CDMS has excluded some
points with large $\lambda$ in the nMSSM, there are still many
surviving points with $\lambda$ as large as 0.7.

In Fig.~\ref{fig8} we show the decay branching ratio of $h^0 \to
\chi^0_1 \chi^0_1$ versus the mass of the SM-like Higgs boson $h^0$.
Such a decay is strongly correlated to the $\chi$-nucleon scattering
because the coupling $h^0 \chi^0_1\chi^0_1$ is involved in both
processes. We see that in the MSSM and NMSSM this decay mode can
open only in a very narrow parameter space since $\chi_1^0$ cannot
be so light, and in the allowed region this decay has a very small
branching ratio (below $10\%$). By contrast, in the nMSSM this decay
can open in a large part of the parameter space since the LSP can be
very light, and its branching ratio can be quite large (over $80\%$
or  $90\%$). Such a large invisible decay ratio may indicate a
severe challenge for finding the Higgs boson $h^0$ at the LHC if the
nMSSM is the true story. Fig.~\ref{fig8} also indicates that the mass
of $h^0$ can reach $160$ GeV. We checked that these cases correspond
to $\lambda$ varying from 0.6 to 0.7 so that the mass is enhanced at
tree level.

Now we discuss the reason for the peculiarity of the nMSSM
predictions shown from Fig.~\ref{fig2} to Fig.~\ref{fig8}. About the
narrow parameter space of the nMSSM constrained by collider
experiments, a detailed analysis has been given in \cite{xnMSSM2},
here we only explain the behavior of the nMSSM under the
CDMS/SuperCDMS limits. Our explanation is based on following three
facts.  The first comes from the neutralino mass matrix in
Eq.(\ref{mass-matrix3}) which implies that $m_{\chi^0_1}$ can be
written as \cite{xnMSSM2}:
\begin{eqnarray}
m_{\chi_1^0} \simeq \frac{2 \mu \lambda^2 ( v_u^2 + v_d^2
)}{2 \mu^2 + \lambda^2 (v_u^2 + v_d^2 )} \frac{\tan \beta}{\tan^2
\beta + 1 }.\label{m_chi0}
\end{eqnarray}
This formula shows that to get a heavy $\chi^0_1$, we need a
large $\lambda$, a small $\tan \beta$ as well as a moderate $\mu$.
The second fact is that,  due to the singlino dominance of $\chi^0_1$ in the
nMSSM, the interaction of $\chi^0_1$ with squarks is suppressed
and the Higgs mediated contribution in Figs.\ref{Feyn1} and \ref{Feyn2}
then becomes dominant in the scattering.  In this case, $\lambda$
determines the size of the scattering for a given $m_{\chi^0_1}$ by
affecting the coupling $T_{h00}$ \cite{NMSSMTools} and a large
$\lambda$ can enhance the cross section.  The last fact is based on
Fig.~\ref{fig1} which shows that the constraints of CDMS results
become stringent for heavy $\tilde{\chi}^0_1$ and as a result, only
$m_{\tilde{\chi}_1^0}$ around 40 GeV  is excluded by CDMS. With
these facts, one can easily understand the features of
Figs.~\ref{fig2}-\ref{fig8}. For example, Fig.~\ref{fig6} and
Fig.~\ref{fig7} indicate that the disfavored points by CDMS are
characterized by small $\tan \beta$ and large $\lambda$. The reason
is that only under these two conditions, both $m_{\tilde{\chi}}$
and the cross section can be large simultaneously.

The similarity of the allowed parameter space for the MSSM and NMSSM
can be understood as follows. In both models $\chi^0_1$ is composed
dominantly by bino, as shown in Fig.~\ref{fig2}. Then the properties
of $\chi^0_1$ (like the relic density and the $\chi$-nucleon
scattering) are similar in both models. Our such conclusion
agrees with \cite{Barger} except that the conclusion of
\cite{Barger} is based on a different scan scheme.
Compared with \cite{Barger}, we considered more constraints and so
our conclusions are more robust.

\section{summary}
\label{sec4}
Considering the current direct and indirect collider constraints,
we gave a comparative study for the neutralino dark matter scattering on nucleon
in the MSSM, the NMSSM and the nMSSM.
We showed the predictions for the elastic cross section by scanning over the
parameter space allowed by the collider constraints
and demonstrated the property of the allowed parameter space with/without
the new limits from CDMS II. We found that for each model the new CDMS limits can
exclude a large part of parameter space allowed by current collider constraints.
The property of the allowed parameter space is found to be similar for MSSM and NMSSM,
but quite different for nMSSM.
Further, the future SuperCDMS can cover most part of the allowed  parameter space
for each model.

\section*{Acknowledgment}
This work was supported in part by HASTIT under grant No. 2009HASTIT004,
by the Grant-in-Aid for Scientific Research (No. 14046201) from the Japan
Ministry of Education, Culture, Sports, Science and Technology,
by the National Natural Science Foundation of China (NNSFC) under grant
Nos. 10505007, 10821504, 10725526 and 10635030,
and by the Project of Knowledge
Innovation Program (PKIP) of Chinese Academy of Sciences under
grant No. KJCX2.YW.W10.

\appendix

\section{Spin-Independent Cross section of $\chi$-nucleon scattering}

\begin{figure}[htbp]
 \epsfig{file=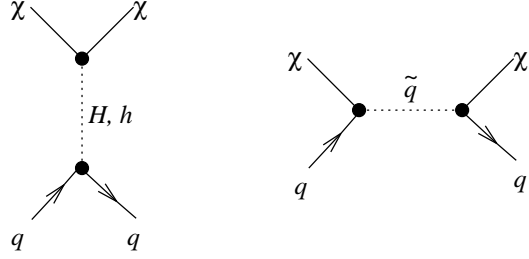,height=4cm}
 \vspace{-0.3cm}
\caption{Feynman diagrams contributing to the scalar
elastic-scattering amplitude of a neutralino
    from quarks in the MSSM, where $H$ and $h$ denote the CP-even Higgs bosons and $\tilde{q}$ represents a scalar quark. }
   \label{Feyn1}
    \end{figure}
\begin{figure}[htbp]
 \epsfig{file=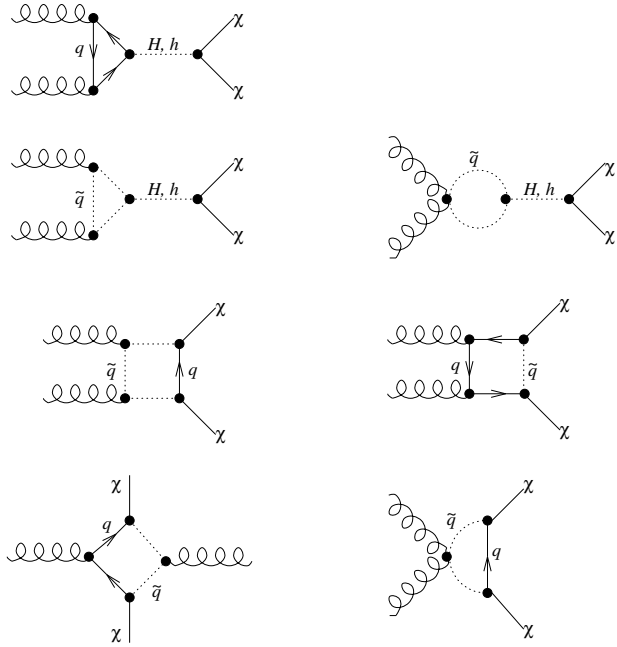,height=8.5cm}
 \vspace{-0.3cm}
\caption{Feynman diagrams contributing to the gluonic interaction
    with neutralinos, which contributes to the scalar
    elastic-scattering amplitude for neutralinos and nuclei.}
   \label{Feyn2}
    \end{figure}

In supersymmetric models, the spin-independent elastic
$\chi$-nucleon scattering is described by the following
effective Lagrangian\cite{Drees,susy-dm-review}:
\begin{eqnarray}
    {\cal L} & = & f_q \bar{\chi} \chi \bar{q} q
          + g_q \left[ -2 i \bar{\chi} \gamma^\mu \partial^\nu \chi
                {\cal O}^{(2)}_{q \mu\nu}
            -{1\over 2} m_q m_\chi \bar{q} q \bar{\chi}\chi
        \right ]  \nonumber  \\
    & & + b\, \alpha_s \bar{\chi} \chi G^a_{\mu\nu} G^{a\mu\nu}
        - \alpha_s(B_{1D} + B_{1S})
            \bar{\chi}\partial_\mu \partial_\nu \chi\;
            {\cal G}^{(2)\mu\nu}  \nonumber  \\
    & & + \alpha_s B_{2S} \bar{\chi}
            \left( i \partial_\mu \gamma_\nu
                + i \partial_\nu \gamma_\mu \right) \chi\;
            {\cal G}^{(2)\mu\nu},  \label{EFT}
\end{eqnarray}
where the twist-two quark and gluon operators are defined by
\begin{eqnarray}
    {\cal O}^{(2)}_{q\mu\nu} & = & {i\over 2}
        \left[ \bar{q} \gamma_\mu \partial_\nu q
        + \bar{q} \gamma_\nu \partial_\mu q
        - {1\over 2} g_{\mu\nu} \bar{q} \not{\hspace*{-0.08cm} \partial} q
        \right], \nonumber \\
    {\cal G}^{(2)\mu\nu} & = & G^{a\mu}_\rho G^{a\rho\nu}
        + {1\over 4} g^{\mu\nu} G^{a\sigma\rho} G^a_{\sigma\rho},
\end{eqnarray}
$G_a^{\mu\nu}$ is the gluon field-strength tensor, and $f_q$, $g_q$, $b$ and $B$ are coefficients.

In the MSSM, the coefficients $f_q$ and $q_q$ are determined by calculating the diagrams in
Fig.~\ref{Feyn1} in the extreme nonrelativistic limit, and they are
given by
\begin{eqnarray}
    f_{q_i} &=& f_{q_i}^{\tilde{q}} + f_{q_i}^H = -{1\over 4} \sum_{\tilde{q}_j}
        {X_{q\; ij\, 0}^\prime W_{q\; ij\, 0}^\prime
            \over
        m_{\tilde{q}_j}^2 - (m_\chi + m_{q_i})^2 } \;\; + \sum_{h={h^0, H^0}}
        {g T_{h00} T_{h q_i q_i} \over 2 m_h^2}, \nonumber \\
    g_{q_i} &= & -{1\over 8} \sum_{\tilde{q}_j}
        { \left(X_{q\; ij\, 0}^\prime\right)^2
          + \left( W_{q\; ij\, 0}^\prime\right)^2
            \over
        \left[m_{\tilde{q}_j}^2 - (m_\chi + m_{q_i})^2 \right]^2 },
        \label{treecontr}
\end{eqnarray}
where the subscripts $q=u,d$ and $i=1,2,3$ refers to the flavor
index in quark sector, and $X^\prime_{qij0}$, $W^\prime_{qij0}$,
$T_{h00}$ and $T_{h q_i q_i}$ are the coupling coefficients of
$\bar{q}_i P_R \chi_0 \tilde{q}_j$, $\bar{q}_i P_R \chi_0
\tilde{q}_j$, $\bar{\chi}_0\chi_0 h$ and $\bar{q}_iq_i h$ vertices
respectively.

The coefficients of the last four operators can be obtained in a
similar way from Fig.~\ref{Feyn2}, and their expressions are
\begin{eqnarray}
b &=& - T_{\tilde{q}} + B_D + B_S - {m_\chi\over 2} B_{2S}
        - {m_\chi^2 \over 4} (B_{1D} + B_{1S}), \nonumber \\
T_{\tilde q} & = & {1\over 96 \pi} \sum_{h=h^0,H^0}
            {g\over 2} {T_{h\, 00} \over m_h^2}
            \sum_{\tilde{q}_j}
            {g_{h {\tilde q}_j {\tilde q}_j}
            \over m_{{\tilde q}_j}^2}  \nonumber\\
B_{D} & = &{1\over 32 \pi} \sum_{q_i, \tilde{q}_j} m_{q_i}
            X_{q\; ij\, 0}^\prime W_{q\; ij\, 0}^\prime
            \; I_1(m_{q_i}, m_{\tilde{q}_j}, m_\chi), \nonumber \nonumber \\
B_{S} & = & {1\over 32 \pi} \sum_{q_i, \tilde{q}_j} m_\chi
            {1\over 2} \left[
            \left( X_{q\; ij\, 0}^\prime \right)^2 +
            \left( W_{q\; ij\, 0}^\prime \right)^2
            \right]
            \; I_2(m_{q_i}, m_{\tilde{q}_j}, m_\chi), \nonumber \\
B_{1D} & = & {1\over 12 \pi} \sum_{q_i, \tilde{q}_j} m_{q_i}
            X_{q\; ij\, 0}^\prime W_{q\; ij\, 0}^\prime
            \; I_3(m_{q_i}, m_{\tilde{q}_j}, m_\chi), \nonumber \\
B_{1S} & = & {1\over 12 \pi} \sum_{q_i, \tilde{q}_j} m_\chi
            {1\over 2} \left[
            \left( X_{q\; ij\, 0}^\prime \right)^2 +
            \left( W_{q\; ij\, 0}^\prime \right)^2
            \right]
            \; I_4(m_{q_i}, m_{\tilde{q}_j}, m_\chi),  \nonumber \\
B_{2S} & = & {1\over 48 \pi} \sum_{q_i, \tilde{q}_j}
            {1\over 2} \left[
            \left( X_{q\; ij\, 0}^\prime \right)^2 +
            \left( W_{q\; ij\, 0}^\prime \right)^2
            \right]
            \; I_5(m_{q_i}, m_{\tilde{q}_j}, m_\chi),
            \label{coef1}
\end{eqnarray}
where $I_k$s are functions given in Eqs. (B.1a-e) of \cite{Drees} with
the Eq. (B.1d) corrected as follows: the factor in
the first term should read $(m_{\tilde q}^2 - m_q^2 - m_\chi^2)$,
with a corrected exponent for $m_\chi$; the term immediately
following should read $-1/ m_{\tilde q}^2 m_\chi^4$, again with a
corrected exponent for $m_\chi$; finally, a sign in the last term
should be corrected so that it reads $\left[\cdots - m_{\tilde q}^2
+ m_\chi^2 \right] L$.

About above formulae, two points should be noted. One is the Lagrangian
in Eq.(\ref{EFT}) is specified at a  high-energy scale, for example,
$\mu_0\simeq m_h $, and in order to get the scattering rate measured in
dark matter direct detection experiments, one must consider important QCD and
SUSY-QCD corrections to the coefficients\cite{Djoudi}. In our calculation, we have
considered such effect. The other is some extensions of the MSSM,
such as NMSSM and nMSSM considered in this paper, usually predict extra
CP-even Higgs bosons and neutralinos, and consequently,  the couplings
appeared in above formulae may be changed. In this case, the
formulae listed above still keep valid in the sense that one must use
the corresponding new couplings with the same convention as that in\cite{susy-dm-review} and
also include the contributions from new intermediate states.
For example, the NMSSM predicts three CP-even Higgs bosons, and one should
add the three boson contributions in getting $f_{q_i}^H$\cite{Bednyakov}.

Given the effective Lagrangian in Eq.(\ref{EFT}), one can write down the
spin-independent scattering cross section of a neutralino from a nucleon N
(proton or neutron) in a standard way\cite{Drees,Belanger}:
\begin{eqnarray}
\sigma^{SI} = \frac{4 m_r^2}{\pi} f_N^2
\end{eqnarray}
where $m_r = \frac{m_\chi m_N}{ m_\chi + m_N }$ is the reduced LSP
mass, and $f_N$ is the effective couplings of the neutralino to nucleon,
which is given by:
\begin{eqnarray}
    {f_N \over m_N} &=& \sum_{q=u,d,s}
        {f_{Tq}^N \over m_q}
        \left[ f_q - {m_\chi m_{q} \over 2} g_q \right]
        + {2\over 27} f_{TG}^N \sum_{q=c,b,t} {f_q^{H}
        \over m_q} \nonumber \\
        & & - {3\over 2} m_\chi \sum_{q=u,d,s,c,b }
        g_q(\mu_0) q^N(\mu_0^2) - {8\pi \over 9}\, b\,  f_{TG}^N
        \nonumber \\
        & & + {3\over 2} m_\chi G^N (\mu_0^2) \alpha_s(\mu_0^2)
            \left[ B_{2S} + {m_\chi\over 2}
                 \left( B_{1D}+B_{1S} \right) \right].  \label{effcoup}
\end{eqnarray}
In Eq.(\ref{effcoup}), $f_{Tq}^N$ denotes the
fraction of the nucleon mass $m_N$ that is due to the light quark
$q$, and $f_{TG}^N=\frac{2}{27}( 1 - f_{Tu}^N - f_{Td}^N - f_{Ts}^N )$
is the heavy quark contribution to $m_N$,  which is induced via gluon exchange.
The function $q^N (\mu_0^2)$ and
$G^N (\mu_0^2)$ appear in the second moment of the
quark (gluon) distribution functions and they represent  the quark and
gluon densities in the nucleon at the scale $\mu_0$. The quantities $g_q(\mu_0)
q^N (\mu_0^2)$ and $G^N (\mu_0^2)\alpha_s(\mu_0^2)$ in the
third term and the last term is a renormalization-group invariant
(in other words, independent of $\mu_0$) and their evaluation was
described in detail in \cite{Drees}. In our calculation, we use $\sigma_{\pi
N}=64$ MeV and $\sigma_0 = 35$ MeV to get the values of $f_{Tq}^N$ and
use CTEQ6L to get the values of $q^N (m_b^2)$ and $G^N(m_t^2)$.

Before we end this section, we remind two subtleties in
Eq.(\ref{effcoup})\cite{Drees}. One is to get the coefficient $b$
by the formula in Eq.(\ref{coef1}), one should not include the
contribution of $u,d,s$ quarks to $B_D$ since they are
non-perturbative effects. The other is only top quark contribution
needs to be considered in getting $B_{2S}$ in the last term. The
reason is the contributions from $u,d,s$ quarks to $B_{2S}$ are
non-perturbative effects, and the contributions from $c,b$ quarks
have been moved to the third term of Eq.(\ref{effcoup}).


\begin{thebibliography}{99}
\bibitem{mssm} For a review, see, e.g.,
  H.~E.~Haber and G.~L.~Kane,
  Phys.\ Rept.\  {\bf 117}, 75 (1985).
\bibitem{NMSSM} See, e.g.,
   J.~R.~Ellis,  {\it{et al.}}, \PRD39, 844 (1989);
   M.~Drees, Int.\ J.\ Mod.\ Phys.\ A {\bf 4}, 3635 (1989).
   S.~F.~King, P.~L.~White, \PRD52, 4183 (1995);
   B. Ananthanarayan, P.N. Pandita, \PLB353, 70 (1995); \PLB371, 245 (1996);
                                    Int. J. Mod. Phys. A12, 2321 (1997);
   B. A. Dobrescu, K. T. Matchev, \JHEP0009, 031 (2000);
   V. Barger, P. Langacker, H.-S. Lee, G. Shaughnessy, \PRD73,(2006) 115010;
    R.~Dermisek, J.~F.~Gunion, \PRL95, 041801 (2005);
    G.~Hiller, \PRD70, 034018 (2004);
    F.~Domingo, U.~Ellwanger, \JHEP0712, 090 (2007);
    Z.~Heng,  {\it et al.}, \PRD77, 095012 (2008);
    R. N. Hodgkinson, A. Pilaftsis, \PRD76, 015007 (2007); \PRD78, 075004 (2008);
    W. Wang, Z. Xiong, J. M. Yang, \PLB680, 167 (2009);
    J. Cao, J. M. Yang, \JHEP0812, 006 (2008); \PRD78, 115001 (2008);

 \bibitem{xnMSSM1}
  P. Fayet, \NPB90, 104 (1975);
  C.~Panagiotakopoulos, K.~Tamvakis,
   \PLB446, 224 (1999); \PLB469, 145 (1999);
  C.~Panagiotakopoulos, A. Pilaftsis, \PRD63, 055003 (2001);
  A.~Dedes, {\it et al.}, \PRD63, 055009 (2001);
  A.~Menon, {\it et al.}, \PRD70, 035005 (2004);
  V.~Barger, {\it et al.}, \PLB630, 85 (2005).
  C.~Balazs, {\it et al.}, \JHEP0706, 066 (2007).

\bibitem{xnMSSM2}
  J. Cao, H. E. Logan, J. M. Yang, \PRD79, 091701 (2009).

\bibitem{hooper} D.~Hooper and T.~M.~P.~Tait, \PRD80, 055028 (2009);
                 W. Wang, {\it et al.}, \JHEP0911, 053 (2009);
                 Y.~Bai, M.~Carena, J.~Lykken, arXiv:0905.2964.

\bibitem{pamela} O.~Adriani {\it et al.}, PAMELA Collaboration,
                  Nature {\bf 458}, 607 (2009).

\bibitem{string}  O. Lebedev and S. Ramos-Sanchez, arXiv:0912.0477.

\bibitem{cdms2} Z. Ahmed, {\it et al.}, CDMS-II Collaboration,
                arXiv:0912.3592.

\bibitem{cdms2-a} For recent works motivated by CDMS II results, see, e.g.,
   M.~Kadastik, K.~Kannike, A.~Racioppi and M.~Raidal, arXiv:0912.3797;
  N.~Bernal and A.~Goudelis, arXiv:0912.3905;
  A.~Bottino, F.~Donato, N.~Fornengo and S.~Scopel, arXiv:0912.4025;
  D.~Feldman, Z.~Liu and P.~Nath, arXiv:0912.4217;
  M.~Ibe and T.~T.~Yanagida, arXiv:0912.4221;
  R.~Allahverdi, B.~Dutta and Y.~Santoso, arXiv:0912.4329;
  M.~Endo, S.~Shirai and K.~Yonekura, arXiv:0912.4484;
  Q.~H.~Cao, I.~Low and G.~Shaughnessy, arXiv:0912.4510;
  Q.~H.~Cao, C.~R.~Chen, C.~S.~Li and H.~Zhang, arXiv:0912.4511;
  K.~Cheung and T.~C.~Yuan, arXiv:0912.4599;
  J.~Hisano, K.~Nakayama and M.~Yamanaka, arXiv:0912.4701;
  X.~G.~He, {\it et al.}, arXiv:0912.4722;
  I.~Gogoladze, R.~Khalid, S.~Raza and Q.~Shafi, arXiv:0912.5411;
  M.~Aoki, S.~Kanemura and O.~Seto, arXiv:0912.5536;
  R.~Foot, arXiv:1001.0096;
  M.~Asano and R.~Kitano, arXiv:1001.0486;
  W.~S.~Cho et al., arXiv:1001.0579;
  J. Shu, P. F. Yi, S. H. Zhu, arXiv:1001.1076;
  D. P. Roy, arXiv:1001.4346;
  S. Khalil, H. S. Lee, E. Ma, arXiv:1002.0692;
  A. Bandyopadhyay, {\it et al.}, arXiv:1002.0753; arXiv:1003.0809;
  J. Hisano, {\it et al.}, arXiv:1003.3648;
  L. Wang, J. M. Yang, arXiv:1003.4492.

\bibitem{ellis} J. Ellis, A. Ferstl and K. A. Olive, \PLB481, 304 (2000);
                J. Ellis, K. A. Olive, Y. Santoso, V. C. Spanos, \PRD71, 095007 (2005);
                A. Bottino, {\it et al.}, \PLB402, 113 (1997).

\bibitem{supercdms} R. Gaitskell, V. Mandic, and J. Filippini,
                    http://dmtools.berkeley.edu/limitplots.

\bibitem{Yao} C. Amsler, {\it et al.} (Particle Data Group), \PLB667, 1 (2008). 


\bibitem{Higgs} S.~Schael, {\it et al.}, \EPJC47, 547 (2006).

\bibitem{NMSSMTools}
     U.~Ellwanger, J.~F.~Gunion and C.~Hugonie, \JHEP0502, 066 (2005);
     U.~Ellwanger and C.~Hugonie, Comput.\ Phys.\ Commun.\  {\bf 175}, 290 (2006).

\bibitem{Altarelli}
  G.~Altarelli and R.~Barbieri, \PLB253, 161 (1991);
  M. E. Peskin, T. Takeuchi, \PRD46, 381 (1992).

\bibitem{Davier}
  M.~Davier,  {\it et al.}, \EPJC66, 1 (2010).


\bibitem{dmconstr}
C.~L.~Bennett {\it et al.}, Astrophys.\ J.\ Suppl.\ {\bf 148} (2003) 1;
D.~N.~Spergel {\it et al.}, Astrophys.\ J.\ Suppl.\ {\bf 148} (2003) 175.

\bibitem{zerwas}   D.J. Miller, R. Nevzorov, P.M. Zerwas,  \NPB681, 3 (2004).

\bibitem{susy-dm-review} G. Junman, M. Kamionkowski and K. Griest,
                         Phys. Rept. 267, 195 (1996).
\bibitem{Drees}
  M.~Drees and M.~Nojiri, \PRD48, 3483 (1993).

\bibitem{Barger}
  V.~Barger, {\it et al.},
  Phys.\ Rev.\  D {\bf 75}, 115002 (2007).

\bibitem{Djoudi}   A.~Djouadi and M.~Drees,
  Phys.\ Lett.\  B {\bf 484}, 183 (2000).

\bibitem{Bednyakov}
  V.~A.~Bednyakov and H.~V.~Klapdor-Kleingrothaus, \PRD59, 023514 (1999);  
  D.~G.~Cerdeno, {\it et al.}, \JHEP0412, 048 (2004);   
  G.~Belanger, C.~Hugonie and A.~Pukhov, JCAP {\bf 0901}, 023 (2009).

\bibitem{Belanger}
  G.~Belanger, F.~Boudjema, A.~Pukhov and A.~Semenov,
  Comput.\ Phys.\ Commun.\  {\bf 180}, 747 (2009).

\end{thebibliography}
\end{document}